\documentclass[preprint,tightenlines,amsmath,amssymb,nofootinbib]{revtex4}
\usepackage{bm} 
\usepackage{hyperref} 

\newcommand{\qt}{\tilde{q}}
\newcommand{\rvec}{\bm{r}}
\newcommand{\ovec}{\bm{\omega}}
\newcommand{\Ovec}{\bm{\Omega}}
\newcommand{\dmu}{\partial_{\mu}}
\newcommand{\dumu}{\partial^{\mu}}
\newcommand{\dnu}{\partial_{\nu}}

\newcommand{\eumn}{\epsilon^{\mu \nu}}
\newcommand{\eqsp}{\mbox{}=\mbox{}}

\newcommand{\Reals}{\mathbb{R}}
\newcommand{\orderO}{\mathcal{O}}
\newcommand{\calN}{\mathcal{N}}
\newcommand{\tp}{\theta^{+}}
\newcommand{\tm}{\theta^{-}}
\newcommand{\tpb}{\bar{\theta}^{+}}
\newcommand{\tmb}{\bar{\theta}^{-}}
\newcommand{\sqt}{\sqrt{2}\,} 
\newcommand{\Qt}{\tilde{Q}}

\newcommand{\form}[1]{\mathsf{#1}}

\newcommand{\chisigchi}[2]{(\chi\chi)^{#1}_{#2}}

\newcommand{\Ctens}[3]{{#1}^{#2}{}_{#3}}
\newcommand{\Tors}[2]{\Ctens{T}{#1}{#2}}
\newcommand{\Chris}[2]{\Ctens{\Gamma}{#1}{#2}}

\begin{document}

\preprint{hep-th/0507204}
\preprint{EFI-05-05}

\title{Worldsheet Instanton Corrections to the Kaluza-Klein Monopole}

\author{Jeffrey A. Harvey}
\email{harvey@theory.uchicago.edu}

\author{Steuard Jensen}
\email{sjensen@theory.uchicago.edu}

\affiliation{Enrico Fermi Institute and Department of Physics,%
  University of Chicago,\\%
  5640 S. Ellis Avenue, Chicago IL 60637, USA}

\date{July 20, 2005}

\begin{abstract}
The Kaluza-Klein monopole is a well known object in both gravity and
string theory, related by T-duality to a ``smeared'' NS5-brane which
retains the isometry around the duality circle.  As the true NS5-brane
solution is localized at a point on the circle, duality implies that
the Kaluza-Klein monopole should show some corresponding behavior.  In
this paper, we express the Kaluza-Klein monopole as a gauged linear
sigma model in two dimensions and show that worldsheet instantons give
corrections to its geometry.  These corrections can be understood as a
localization in ``winding space'' which could be probed by strings
with winding charge around the circle.
\end{abstract}

\maketitle

\section{Introduction}

It is well known that superstring theory compactified on a circle
contains Kaluza-Klein monopoles~\cite{Sorkin:1983ns,Gross:1983hb},
which have an isometry around the circle.  T-duality in that direction
transforms them into $H$-monopoles~\cite{Banks:1988rj,Ooguri:1995wj},
which are understood in string theory as
NS5-branes~\cite{Gauntlett:1992nn,Khuri:1992ww}.  This relationship
forms an important part of the duality web.

However, there are two serious gaps in this familiar story.  The first
is that NS5-branes naturally correspond to the localized $H$-monopole
geometry found in~\cite{Gauntlett:1992nn}.  But that solution breaks
the isometry around the circle and has a throat behavior at short
distances (at least when the NS5-brane charge is greater than one).
This is qualitatively very different from the Kaluza-Klein monopole
solution, and thus would seem to conflict with the basic premise of
T-duality that the physics of the dual solutions is the same.

On this basis, Gregory, Moore, and one of the
authors~\cite{Gregory:1997te} argued that the proper Kaluza-Klein
monopole solution in string theory should be modified.  They suggested
that classical values for string winding states near the monopole core
would lead to a ``throat'' that could be probed by scattering winding
strings.  However, they did not find the corrected geometry
explicitly, and the mechanism generating the necessary corrections
remained unknown.

The second gap in the story is that the natural T-dual of a
Kaluza-Klein monopole geometry is a \emph{smeared} $H$-monopole
solution which retains the $S^1$ isometry.  We know, however, that one
of the $H$-monopole's moduli is its position on $S^1$.  There must
therefore be a localized solution, or else changing the $S^1$ location
would not be a true physical modulus as it would not lead to a new
point in the physical configuration space.  This puzzle was resolved
by Tong~\cite{Tong:2002rq}, who demonstrated that worldsheet
instantons in the smeared $H$-monopole background correct the geometry
to reproduce the localized solution of~\cite{Gauntlett:1992nn}.  The
solution to this second problem provides the necessary tools to solve
the first.

In this paper, we show that the unit charge Kaluza-Klein monopole
receives similar corrections from worldsheet instantons.  As its
topology does not admit holomorphic instantons, we begin from an
$\calN = (4,4)$ supersymmetric gauged linear sigma model and study the
``point-like instantons'' described by Witten~\cite{Witten:1993yc}.
Although we are only able to find the corrected solution in a limit,
our results strongly suggest that the Kaluza-Klein monopole in string
theory is localized in ``winding space'' as expected from T-duality.
Our perspective on the corrections differs slightly from that
of~\cite{Gregory:1997te}: while they carry the same conserved charge
as string winding states, there is no direct identification between
the two.

The structure of this paper is as follows.  In
section~\ref{sec:MonopoleReview} we briefly review the relevant
monopole geometries.  In section~\ref{sec:SigmaModels} we state the
supersymmetric gauged linear sigma models describing the Kaluza-Klein
and $H$-monopoles and show that the former reduces to the expected
nonlinear sigma model in the low-energy limit.  In
section~\ref{sec:Instantons} we identify worldsheet instanton
configurations in the gauged linear sigma model and determine their
leading order effect on the geometry.  Finally, in
section~\ref{sec:InterpretConclude} we relate these results to the
expected properties of the solution and conclude.

\section{Review of monopole geometries}
\label{sec:MonopoleReview}

The usual Kaluza-Klein monopole metric is that of Taub-NUT space, with
no excitation of the antisymmetric tensor or dilaton.  In our
conventions,
\begin{equation}
\label{eq:KKmonmetric}
ds^2 = H(r)\, d \rvec \cdot d \rvec + H(r)^{-1}
 \left( d \kappa + \tfrac{1}{2} \ovec \cdot d \rvec \right)^2\quad.
\end{equation}
Here, $\rvec$ is a position in $\Reals^3$, $\ovec$ is a vector in
$\Reals^3$ satisfying $\nabla \times \ovec = -2 \nabla H(r) =
 -\nabla (1/r)$, and $\kappa$ has period $2 \pi$.  The harmonic
function $H(r)$ is
\begin{equation*}
H(r) = \frac{1}{g^2} + \frac{1}{2 r}\quad,
\end{equation*}
making $g$ the asymptotic radius of the $\kappa$ circle (in string
units).\footnote{To compare with the more common conventions
of~\cite{Gregory:1997te}, the radius is $g = R / \sqrt{\alpha'}$, the
coordinates are $r_{\text{here}} = r_{\text{there}} \cdot R /
\alpha'$ and $\kappa = x^5 / R$, the metric is $ds_{\text{here}}^2 =
-ds_{\text{there}}^2 / \alpha'$, and the harmonic function in the
metric there is $U^{-1} = g^2 H = 1 + R/2r_{\text{there}}$.}  This
solution approaches flat space at the origin and has global topology
$\Reals^4$: its local $\Reals^3 \times S^1$ structure is simply the
Hopf fibration of $S^1$ over $S^2 (\times \Reals_+)$.  After
dimensional reduction on the $S^1$, $\ovec$ gives the vector potential
of a magnetic monopole for the Kaluza-Klein gauge field.  In spherical
coordinates $\{r,\vartheta,\varphi\}$ on $\Reals^3$, one common gauge
choice gives $\omega_r = \omega_\vartheta = 0$, $\omega_\varphi = 1 -
\cos \vartheta$.

In addition to its three collective coordinates corresponding to
position in $\Reals^3$, the Kaluza-Klein monopole has a fourth
collective coordinate related to the antisymmetric 2-form
$B_{mn}$~\cite{Sen:1997zb}.  This dyonic coordinate $\beta$ arises
from the harmonic 2-form of Taub-NUT space, and can be found as a
large gauge transformation of $B_{mn}$:
\begin{equation}
\label{eq:dyoncoord}
\form{B} = \beta\, d \form{\Lambda}\;,
\quad\text{for}\quad
\form{\Lambda} = \frac{1}{g^2 H(r)}
 \left( d\kappa + \tfrac{1}{2} \ovec \cdot d \rvec \right)
\quad.
\end{equation}
Although it is pure gauge, this $\form{B}$ is physically significant
because $\Lambda$ does not vanish at infinity.  If $\beta$ is
constant, $\form{B}$ is a closed form and has no effect on the
geometry, but (for instance) a time-varying $\beta$ carries string
winding charge, which corresponds after dimensional reduction to
electric charge under $B_{m4}$ (where $r^4 \equiv \kappa$).

The (smeared) $H$-monopole solution can be found from this by the
usual Buscher rules for T-duality~\cite{Buscher:1987qj}.  Applying
them to the Kaluza-Klein monopole (with $\beta=0$) gives a solution in
terms of the dual coordinate $r^4 \equiv \theta$:
\begin{equation*}
ds^2 = H(r) \left( d\rvec \cdot d\rvec + d\theta^2 \right)\;,
\qquad
B_{m4} = -\ovec/2
\quad (m=1,2,3)
\quad,
\end{equation*}
with all other independent $B_{mn}$ zero.  (The dilaton becomes
$e^\Phi = H(r)$.)  The physical contribution of $B_{mn}$ is the
torsion $\form{T}=-\form{H}=-d\form{B}$, which can be written in terms
of $H(r)$: $H_{mnp} = \epsilon_{mnp}{}^q H^{-1} \partial_q H$.
Applying the Buscher rules to the general solution with $\beta \neq 0$
reduces to the same form after a coordinate transformation $\theta \to
\theta - \beta/(g^2 H(r))$ and a gauge transformation of $B_{mn}$
(with a gauge parameter that does vanish at infinity).  Thus, up to a
coordinate transformation that is trivial at infinity, $\beta$ simply
corresponds to a shift in $\theta$.

Finally, the localized $H$-monopole can be constructed from a periodic
array of NS5-branes~\cite{Gauntlett:1992nn}.  The forms of the metric
and torsion are the same as in the smeared case above, but the
harmonic function is modified:
\begin{equation}
\label{eq:Hlocalized}
H(r,\theta) = \frac{1}{g^2} + \frac{1}{2r}
 \frac{\sinh r}{\cosh r - \cos \theta}
 =  \frac{1}{g^2} + \frac{1}{2r}
  \sum_{k=-\infty}^\infty e^{-|k|r+ik\theta}
\quad.
\end{equation}
The Fourier-expanded form is directly related to the instanton sum.
In these coordinates the monopole is localized at $\theta=0$; more
generally, we can introduce a constant offset $(\theta - \theta_0)$.

\section{Sigma model actions for monopoles}
\label{sec:SigmaModels}

To study these monopole configurations in string theory, our first
step must be to identify the appropriate worldsheet theories
describing them.  These geometries contain no holomorphic curves (at
least for unit monopole charge), so we will consider worldsheet
instanton effects due to point-like instantons in gauged linear sigma
models~\cite{Witten:1993yc}.  The appropriate models have $\calN =
(4,4)$ supersymmetry in two dimensions.

Our approach in this section generally follows that of
Tong~\cite{Tong:2002rq}, although our conventions and methods differ
slightly.  We introduce the T-dual monopole actions in $\calN = (2,2)$
superspace after defining the necessary superfields.  We then expand
the actions in components as will be necessary for the instanton
calculation in section~\ref{sec:Instantons}.  To make contact with the
Kaluza-Klein geometry given above, we then take the low energy limit
and show that it reduces to the expected nonlinear sigma model.  This
limit will be used in interpreting the results of the instanton
calculation.

\subsection{Superfield definitions}

The gauged linear sigma model actions for the $H$-monopole and the
Kaluza-Klein monopole are constructed from $\calN=(4,4)$
supermultiplets in two dimensions.  Each action includes a gauge
multiplet and a charged hypermultiplet, but where the $H$-monopole
also has a twisted hypermultiplet the Kaluza-Klein monopole has a
normal hypermultiplet instead.  For ease of computation, we decompose
each of these $\calN = (4,4)$ supermultiplets into a pair of $\calN =
(2,2)$ superfields.  (See~\cite{Witten:1993yc} for a review of these
models.)

We begin with the $\calN = (4,4)$ vector multiplet, which decomposes
into an $\calN = (2,2)$ chiral superfield $\Phi$ and vector superfield
$V$.  In terms of component fields (with derivative terms suppressed),
these are
\begin{equation*}
\label{eq:vectormult}
\begin{split}
\Phi & = \phi + \sqt \tp \tilde{\lambda}_+ 
        + \sqt \tm \tilde{\lambda}_-
	+ \sqt \tp \tm ( D^1 - i D^2 )
	+ \dotsb \\
V & = \tp \tpb A_+ + \tm \tmb A_-
        - \sqt \tm \tpb \sigma 
	- \sqt \tp \tmb \sigma^\dag\\
    & \quad \mbox{}
        - 2 i \tp \tm \left(\tmb \bar{\lambda}_- 
	                            + \tpb \bar{\lambda}_+ \right)
        - 2 i \tmb \tpb \left(\tp \lambda_+ 
                                      + \tm \lambda_- \right)
	+ 2 \tp \tm \tmb \tpb D^3 \\
\Sigma & \equiv \frac{1}{\sqt\!} \bar{D}_+ D_- V
  = \sigma + i \sqt \tp \bar{\lambda}_+
        - i \sqt \tmb \lambda_-
	+ \sqt \tp \tmb ( D^3 - i F_{01} )
	+ \dotsb \quad.
\end{split}
\end{equation*}
Here, the vector superfield is in Wess-Zumino gauge and we have
defined $A_\pm \equiv A_0 \pm A_1$.  The gauge invariant twisted
chiral superfield $\Sigma$ allows a natural expression for the theta
angle and Fayet-Iliopoulos term and provides a convenient way of
writing the gauge kinetic term.  Following~\cite{Witten:1993yc},
components of fermion doublets are labeled as $\lambda^\alpha =
(\lambda^-, \lambda^+)$ and $\lambda_\alpha = (\lambda_-, \lambda_+)$,
so $\lambda^- = \lambda_+$ and $\lambda^+ = -\lambda_-$.  As usual,
gauge transformations are given by $V \to V + i(\Lambda-\Lambda^\dag)$
for an arbitrary chiral superfield $\Lambda$.

The charged hypermultiplet common to both monopole actions decomposes
into two chiral superfields $Q$ and $\Qt$ with charges $+1$ and $-1$
under the U(1) gauge group, respectively.  Their component expansions
are
\begin{equation*}
\label{eq:hypermult}
\begin{split}
Q & = q + \sqt \tp \psi_+ + \sqt \tm \psi_- + 2 \tp \tm F
        + \dotsb\\
\Qt & = \qt + \sqt \tp \tilde{\psi}_+ + \sqt \tm \tilde{\psi}_- 
        + 2 \tp \tm \tilde{F}
	+ \dotsb \quad.
\end{split}
\end{equation*}

The $H$-monopole action's twisted hypermultiplet can be decomposed
into a chiral superfield $\Psi$ and a twisted chiral superfield
$\Theta$.  Their component expansions are
\begin{equation*}
\label{eq:twistedmult}
\begin{split}
\Psi & = \frac{( r^1 + i r^2 )}{\sqt\!}
           + \sqt \tp \chi_+ + \sqt \tm \chi_- + 2 \tp \tm G
	   + \dotsb\\
\Theta & = \frac{( r^3 + i \theta )}{\sqt\!}
           - i \sqt \tp \bar{\tilde{\chi}}_+
           - i \sqt \tmb \tilde{\chi}_- + 2 \tp \tmb \tilde{G}
	   + \dotsb \quad.
\end{split}
\end{equation*}
Under T-duality, this twisted hypermultiplet is exchanged for the
normal hypermultiplet of the Kaluza-Klein monopole.  The $\Psi$
superfield is unchanged, but its partner is now a second chiral
superfield $\Gamma$ whose component expansion is
\begin{equation}
\label{eq:Gammadefn}
g^2 \Gamma = \frac{( -{r^3} + i g^2 \gamma )}{\sqt\!}
           + i \sqt \tp \bar{\tilde{\chi}}_+
           + i \sqt \tm \bar{\tilde{\chi}}_- + 2 \tp \tm \tilde{G}'
	   + \dotsb \quad.
\end{equation}
Some of the component fields listed here share their names with
components of $\Theta$; these identifications are justified below.

\subsection{Gauged linear sigma models and T-duality}

\subsubsection{The monopole actions in superspace}
\label{sec:ActionTdual}

The gauged linear sigma model action corresponding to the $H$-monopole
written in terms of the above $\calN=(2,2)$ superfields is given by
the sum of the following Lagrangian densities (plus complex conjugates
of the $F$ and $\tilde{F}$ terms):
\begin{equation*}
\label{eq:Hsuperaction}
\begin{gathered}
\mathcal{L}_D = \int d^4\theta \left[
  \frac{1}{e^2} \left( -\Sigma^\dag \Sigma + \Phi^\dag \Phi \right)
  + \frac{1}{g^2} \left( -\Theta^\dag \Theta + \Psi^\dag \Psi \right)
  + Q^\dag e^{2 V} Q + \Qt^\dag e^{-2 V} \Qt \right]\\
\mathcal{L}_F = \int d^2\theta \ 
  \left( \sqt \Qt \Phi Q - \Phi \Psi \right) \hspace{5em}
\mathcal{L}_{\tilde{F}} = - \int d^2\vartheta \ 
\Theta \Sigma \quad.
\end{gathered}
\end{equation*}
Here, $d^2\theta \equiv -d\tp d\tm/2$ and $d^2\vartheta \equiv -d\tp
d\tmb/2$ are the usual measures on chiral and twisted chiral
superspace, respectively.  We begin from the $H$-monopole action in
order to follow the effects of $\mathcal{L}_{\tilde{F}}$ through
T-duality.  As shown below, this term leads to a topologically
significant total derivative of component fields after duality.

To obtain the corresponding superspace action for the Kaluza-Klein
monopole, we find the T-dual action in superspace following Ro\v{c}ek
and Verlinde~\cite{Rocek:1991ps}.  It is first necessary to write the
action uniformly as an integral over full superspace by applying the
identity
\begin{equation}
\label{eq:Ftconvert}
\int d^2\vartheta \ \Sigma \Theta
  + \int d^2\bar{\vartheta} \ \Sigma^\dag \Theta^\dag
  = \int d^4\theta \left[
    \sqt \left( \Theta + \Theta^\dag \right) V \right]
    - \eumn \dmu (\theta A_\nu)\quad.
\end{equation}
We can then write the $\Theta$-dependent part of the action in first
order form, replacing $\Theta + \Theta^\dag$ with a real superfield
$B$ together with a chiral superfield Lagrange multiplier $\Gamma$:
\begin{equation*}
\begin{split}
  \int d^4\theta \left[ -\frac{1}{g^2} \Theta^\dag \Theta 
  - \sqt \left(\Theta + \Theta^\dag \right) V \right]
 & = \int d^4\theta \left[ -\frac{1}{2 g^2}
      \left(\Theta + \Theta^\dag \right)^2
  - \sqt \left(\Theta + \Theta^\dag \right) V \right]\\
 & = \int d^4\theta \left[ -\frac{1}{2 g^2} B^2 - \sqt B V
    - \left(\Gamma + \Gamma^\dag \right) B
  \right] \quad.
\end{split}
\end{equation*}
The final equality can be proved by integrating out $\Gamma$ (and its
conjugate), which requires that we rewrite the $\Gamma$ term as an
integral over chiral superspace only.  Up to total derivatives, $2
\int d^2\bar{\theta}$ is equivalent to $-\bar{D}_+ \bar{D}_-$.  Both
of these derivatives annihilate $\Gamma$ by definition, so the
$\Gamma$ equation of motion is $\bar{D}_+ \bar{D}_- B = 0$.  The
$\Gamma^\dag$ constraint is just the conjugate of this, so we can
write $B = \Theta + \Theta^\dag$ for a twisted chiral superfield
$\Theta$ (for which, by definition, $\bar{D}_+ \Theta = D_- \Theta =
0$).  This combination $\Theta + \Theta^\dag$ has no undifferentiated
imaginary scalar part, so the only source for a constant offset in
$\theta$ in the first order action is the total derivative term from
Eq.~\eqref{eq:Ftconvert}.

To find the T-dual action we instead integrate out $B$, which yields
the equation of motion $B = -g^2 (\Gamma + \Gamma^\dag + \sqt V)$.
This leaves us with the duality substitution
\begin{equation*}
\int d^4\theta \left[ -\frac{1}{g^2} \Theta^\dag \Theta 
  - \sqt \left(\Theta + \Theta^\dag \right) V \right]
 \to
\int d^4\theta\  \frac{g^2}{2}
  \left( \Gamma + \Gamma^\dag + \sqt V \right)^2 \quad.
\end{equation*}
The full superspace action for the Kaluza-Klein monopole is then
constructed from
\begin{gather}
\mathcal{L}_D  = \int d^4\theta \left[
  \frac{1}{e^2} \left( -\Sigma^\dag \Sigma + \Phi^\dag \Phi \right)
  + \frac{g^2}{2}
      \left( \Gamma + \Gamma^\dag + \sqt V \right)^2
 + \frac{1}{g^2} \Psi^\dag \Psi
  + Q^\dag e^{2 V} Q + \Qt^\dag e^{-2 V} \Qt \right] \nonumber \\
\label{eq:Ksuperaction}
\mathcal{L}_F  = \int d^2\theta \ 
  \left( \sqt \Qt \Phi Q - \Phi \Psi \right) \hspace{5em}
\mathcal{L}_{\text{top.}} = \eumn \dmu (\theta A_\nu)
 \quad.
\end{gather}
In order for the action to remain gauge invariant, $\Gamma$ must
transform by a simple shift: $\Gamma \to \Gamma - i \sqt \Lambda$.  In
the case of an ``ordinary'' gauge transformation $\Lambda = \lambda
\in \Reals$ (the residual freedom after fixing Wess-Zumino gauge), the
shift affects only one component: $\gamma \to \gamma - 2 \lambda$.
The total derivative term $\mathcal{L}_{\text{top.}}$ will be
topologically significant in the instanton calculation.

The component fields of $\Theta$ and $\Gamma$ can be related to one
another by equating the two expressions for $B$ found above: $\Theta +
\Theta^\dag = B = -g^2 (\Gamma + \Gamma^\dag + \sqt V)$.  This
justifies the equality between the $\Theta$ and $\Gamma$ component
fields $r^3$ and $\tilde{\chi}_\pm$ asserted in
Eq.~\eqref{eq:Gammadefn}.  The components $\theta$ and $\gamma$ are
not directly related, but their derivatives are: $\partial_\pm
\theta/g = \mp g (\partial_\pm \gamma + A_\pm)$.  The relative $\mp$
is the usual sign change of the right-moving worldsheet coordinate
under T-duality, and the factors of $g$ convert angles to arc lengths.

\subsubsection{The monopole actions in components}

When these superspace actions are expanded in components and the
auxiliary fields are eliminated, the results are almost identical.
Below, we present the results for the Kaluza-Klein monopole; the few
changes required for the $H$-monopole are discussed in the text.

We divide the action into a sum of kinetic, scalar potential,
``Yukawa,'' and topological terms:
\begin{equation*}
\label{eq:totalaction}
S = \frac{1}{2 \pi} \int d^2x \ \left( \mathcal{L}_{\text{kin}}
    + \mathcal{L}_{\text{pot}} + \mathcal{L}_{\text{Yuk}}
    + \mathcal{L}_{\text{top.}} \right)
 \quad.
\end{equation*}
The term $\mathcal{L}_{\text{top.}}$ was defined in
Eq.~\eqref{eq:Ksuperaction}; it is absent from the $H$-monopole
action.  The component form of the kinetic terms is:
\begin{align}
\nonumber
\mathcal{L}_{\text{kin}} & =
     \frac{1}{e^2} \Bigl( \frac{1}{2} F^{2}_{01} 
       - |\dmu \phi|^2 - |\dmu \sigma|^2
       + i (\bar{\lambda}_+ \partial_- \lambda_+ +
            \bar{\tilde{\lambda}}_+ \partial_- \tilde{\lambda}_+ +
	    \bar{\lambda}_- \partial_+ \lambda_- +
            \bar{\tilde{\lambda}}_- \partial_+ \tilde{\lambda}_- )
       \Bigr)\\
\nonumber
 & \quad \mbox{} + \frac{1}{g^2} \Bigl( -\frac{1}{2} |\dmu \rvec|^2
       - \frac{g^4}{2} (\dmu \gamma + A_\mu)^2
       + i (\bar{\chi}_+ \partial_- \chi_+ +
            \bar{\tilde{\chi}}_+ \partial_- \tilde{\chi}_+ +
	    \bar{\chi}_- \partial_+ \chi_- +
            \bar{\tilde{\chi}}_- \partial_+ \tilde{\chi}_- )
       \Bigr)\\
\label{eq:LkinCompts}
 & \quad \mbox{} + \Bigl( 
       - |\mathcal{D}_\mu q|^2 - |\mathcal{D}_\mu \qt|^2
       + i (\bar{\psi}_+ \mathcal{D}_- \psi_+ +
            \bar{\tilde{\psi}}_+ \mathcal{D}_- \tilde{\psi}_+ +
	    \bar{\psi}_- \mathcal{D}_+ \psi_- +
            \bar{\tilde{\psi}}_- \mathcal{D}_+ \tilde{\psi}_- )
       \Bigr) \quad.
\end{align}
The worldsheet metric is flat and given by $\eta_{\mu \nu} = \left(
\begin{smallmatrix} -1&0\\0&1 \end{smallmatrix} \right)$.  We have
defined $\partial_\pm = \partial_0 \pm \partial_1$, and the covariant
derivative is defined as $\mathcal{D}_\mu q = \dmu q + i A_\mu q$ and
$\mathcal{D}_\mu \qt = \dmu \qt - i A_\mu \qt$ (and similarly for
other components from the same supermultiplet).  To obtain the kinetic
terms of the $H$-monopole action, only one substitution is required:
$-\frac{g^4}{2} (\dmu \gamma + A_\mu)^2 \to -\frac{1}{2} (\dmu
\theta)^2$.

After eliminating auxiliary fields, the scalar potential is:
\begin{equation}
\label{eq:LpotCompts}
\begin{split}
\mathcal{L}_{\text{pot}} & =
     - \frac{e^2}{2} \left( |q|^2 - |\qt|^2 - r^3 \right)^2
     - \frac{e^2}{2} \left| 2 \qt q 
       - \left( r^1 + i r^2 \right) \right|^2 \\
 & \quad \mbox{} - \left( |\phi|^2 + |\sigma|^2 \right)
       \left( g^2 + 2 |q|^2 + 2 |\qt|^2 \right)
 \quad.
\end{split}
\end{equation}
The $H$-monopole action also includes the term $-\theta F_{01}$.

Finally, the ``Yukawa'' action (which also includes several
two-fermion terms) is:
\begin{equation*}
\label{eq:LYukCompts}
\begin{split}
\mathcal{L}_{\text{Yuk}} & =
     \left( \tilde{\lambda}_+ \chi_- - \lambda_+ \bar{\tilde{\chi}}_-
     + \bar{\lambda}_+ \tilde{\chi}_- 
     - \bar{\tilde{\lambda}}_+ \bar{\chi}_- \right)
   - \left( \tilde{\lambda}_- \chi_+ - \lambda_- \bar{\tilde{\chi}}_+
     + \bar{\lambda}_- \tilde{\chi}_+ 
     - \bar{\tilde{\lambda}}_- \bar{\chi}_+ \right)\\
 & \quad \mbox{} + i \sqt q \; \left(
       \bar{\lambda}_+ \bar{\psi}_- - \bar{\lambda}_- \bar{\psi}_+
       + i \tilde{\lambda}_+ \tilde{\psi}_- 
       - i \tilde{\lambda}_- \tilde{\psi}_+ \right)
     + \sqt \, \sigma \; \left( \psi_+ \bar{\psi}_-
       - \tilde{\psi}_+ \bar{\tilde{\psi}}_- \right)\\
 & \quad \mbox{} + i \sqt q^\dag \left(
       \lambda_+ \psi_- - \lambda_- \psi_+
       - i \bar{\tilde{\lambda}}_+ \bar{\tilde{\psi}}_- 
       + i \bar{\tilde{\lambda}}_- \bar{\tilde{\psi}}_+ \right)
     - \sqt \sigma^\dag \left( \bar{\psi}_+ \psi_-
       - \bar{\tilde{\psi}}_+ \tilde{\psi}_- \right)\\
 & \quad \mbox{} - i \sqt \qt \; \left(
       \bar{\lambda}_+ \bar{\tilde{\psi}}_- 
       - \bar{\lambda}_- \bar{\tilde{\psi}}_+
       - i \tilde{\lambda}_+ \psi_- 
       + i \tilde{\lambda}_- \psi_+ \right)
     - \sqt \, \phi \; \left( \psi_+ \tilde{\psi}_-
       + \tilde{\psi}_+ \psi_- \right)\\
 & \quad \mbox{} - i \sqt \qt^\dag \left(
       \lambda_+ \tilde{\psi}_- - \lambda_- \tilde{\psi}_+
       + i \bar{\tilde{\lambda}}_+ \bar{\psi}_- 
       - i \bar{\tilde{\lambda}}_- \bar{\psi}_+ \right)
     + \sqt \phi^\dag \left( \bar{\psi}_+ \bar{\tilde{\psi}}_-
       + \bar{\tilde{\psi}}_+ \bar{\psi}_- \right) \quad.
\end{split}
\end{equation*}
The corresponding terms in the $H$-monopole action are identical.

Both actions are invariant under the R-symmetry group $SU(2) \times
SO(4) = [SU(2)]^3$.  The component fields fall into R-multiplets,
which we label by their structure under $SU(2)_V \times SU(2)_L \times
SU(2)_R$.  We define the scalar R-multiplets as $q_i = (q, \qt^\dag)$
(that is, a $(\mathbf{2},\mathbf{1},\mathbf{1})$) and $r^m = (r^1,
r^2, r^3)$ (a $(\mathbf{3},\mathbf{1},\mathbf{1})$).  $\theta$ and
$\gamma$ are R-singlets, and the vector multiplet scalars fall into a
$(\mathbf{1},\mathbf{2},\mathbf{2})$ that obeys a reality condition.
The fermions also fall into R-multiplets: a
$(\mathbf{1},\mathbf{2},\mathbf{1})$ and a
$(\mathbf{1},\mathbf{1},\mathbf{2})$ for the $\psi$s, real multiplets
$(\mathbf{2},\mathbf{\bar{2}},\mathbf{1})$ and
$(\mathbf{2},\mathbf{1},\mathbf{\bar{2}})$ for the $\chi$s, and real
multiplets $(\mathbf{2},\mathbf{2},\mathbf{1})$ and
$(\mathbf{2},\mathbf{1},\mathbf{2})$ for the $\lambda$s.

Finally, a quadratic combination of the $\chi$s into a
$(\mathbf{3},\mathbf{1},\mathbf{1})$ multiplet arises in several
places:
\begin{equation}
\label{eq:chiSqrVector}
\chisigchi{m}{\pm} \equiv
 \Bigl(
  i \left( \chi_\pm \tilde{\chi}_\pm
           + \bar{\chi}_\pm \bar{\tilde{\chi}}_\pm \right),
  \quad
  \left( \chi_\pm \tilde{\chi}_\pm
         - \bar{\chi}_\pm \bar{\tilde{\chi}}_\pm \right),
  \quad
  \left( \bar{\chi}_\pm \chi_\pm
         + \bar{\tilde{\chi}}_\pm \tilde{\chi}_\pm \right)
  \Bigl) \quad.
\end{equation}

\subsection{Low energy limit of the Kaluza-Klein monopole action}

\subsubsection{The low energy limit in superspace}

Nonlinear sigma models for the $H$-monopole and Kaluza-Klein monopole
can be found as the low energy limits of these gauged linear sigma
models, which can be taken in two ways.  The first is based on the
component action as given above.  In our conventions, dimensional
analysis shows that the only dimensionful parameter in this action is
the gauge coupling $e$.  The low energy limit is thus $e^2 \to
\infty$, so the gauge kinetic terms vanish and both of the $q$--$r$
terms in the scalar potential must vanish to ensure finite energy.
The vector multiplet components become auxiliary fields and must be
integrated out, leaving an action which can be written in terms of
twisted hypermultiplet fields alone (after applying the constraints to
eliminate the hypermultiplet fields).  This is essentially the
approach taken by Tong.

The second approach, which we will follow, is to take the $e^2 \to
\infty$ limit while still in the superspace formalism.  The kinetic
terms for $V$ (written in terms of $\Sigma$) and $\Phi$ vanish in
Eq.~\eqref{eq:Ksuperaction}, so both can be treated as auxiliary
superfields and integrated out directly.  The superfield $\Phi$
appears only as a Lagrange multiplier in $\mathcal{L}_F$.  The vector
superfield is somewhat more subtle, as we must deal with the issue of
gauge fixing before integrating it out.  We first restrict to
Wess-Zumino gauge, which fixes all of the gauge freedom except the
ordinary gauge transformations of the component $A_\mu$.  The gauge
choice for this residual symmetry is less crucial; when it is
necessary to make an explicit choice, we will require that $q$ be
purely negative imaginary ($q=-i p$ for real $p>0$).

The vector multiplet equations of motion resulting from this procedure
are
\begin{alignat}{2}
\label{eq:sfieldConstraints}
\Psi & = \sqt \Qt Q
&  \qquad \text{and} \qquad 
\frac{g^2}{\sqt\!} \left( \Gamma + \Gamma^\dag \right)
 = - Q^\dag e^{2 V} Q + \Qt^\dag e^{-2 V} \Qt - g^2 V \quad.
\end{alignat}
It can be verified that these constraints contain precisely the same
information as the vacuum equations and auxiliary field equations of
motion in the component formalism.

When we do go to components, we eventually want to express the full
action in terms of the twisted hypermultiplet fields, but as an
intermediate step, we can (partially) apply the constraints directly
to the Kaluza-Klein monopole action in superspace:
\begin{equation}
\label{eq:KconstrSspaceAction}
\begin{split}
\mathcal{L} & = \int d^4\theta \left[
     \frac{g^2}{2} \left( \Gamma + \Gamma^\dag + \sqt V \right)^2
     + \frac{1}{g^2} \Psi^\dag \Psi
     + Q^\dag e^{2 V} Q + \Qt^\dag e^{-2 V} \Qt \right]\\
 & = \int d^4\theta \left[
     g^2 \Gamma^\dag \Gamma
     + \sqt g^2 V \left( \Gamma + \Gamma^\dag \right) + g^2 V^2
     + \frac{1}{g^2} \Psi^\dag \Psi
     + Q^\dag e^{2 V} Q + \Qt^\dag e^{-2 V} \Qt \right] \\
 & = \int d^4\theta \left[
     g^2 \Gamma^\dag \Gamma - g^2 V^2
     + \frac{1}{g^2} \Psi^\dag \Psi
     + \left( 1 - 2 V^2 \right)
     \left( Q^\dag Q + \Qt^\dag \Qt \right) \right]
\quad.
\end{split}
\end{equation}
In the last line, we have used our choice of Wess-Zumino gauge to
expand the exponentials.  The total derivative term
$\mathcal{L}_{\text{top.}}$ is still present as well, but is not
relevant to these manipulations in superspace.  Completely eliminating
the $Q$ and $V$ superfields at this point would be very difficult due
to the form of the constraints, so the final constraint substitutions
will be carried out at the component level.

\subsubsection{The low energy limit in components}

The nonlinear sigma model for the Kaluza-Klein monopole is found from
this result by expanding Eq.~\eqref{eq:KconstrSspaceAction} in
components and then applying the constraints from
Eq.~\eqref{eq:sfieldConstraints}.  The constraints used to eliminate
the $Q$ and $\Qt$ components in favor of those from $\Psi$ and
$\Gamma$ are
\begin{equation}
\label{eq:comptConstraints}
\begin{aligned}
r^1 + i r^2 = 2 q \qt & \qquad \qquad &
\chi_\pm & = \; \sqt
             \left( \qt \psi_\pm + q \tilde{\psi}_\pm \right) \\
r^3 = |q|^2 - |\qt|^2 & &
\tilde{\chi}_\pm & = i \sqt \left( \qt \bar{\tilde{\psi}}_\pm - 
                                   q \bar{\psi}_\pm \right)
\quad.
\end{aligned}
\end{equation}
These same relationships are found in the $H$-monopole case.  It is
also useful to express $q$ and $\qt$ in terms of the $r^m$, but doing
so is complicated by the fact that together, $q$ and $\qt$ have four
real degrees of freedom while $r^m$ has only three.  We choose to
treat the phase of $q$ as the fourth degree of freedom, so
\begin{equation}
\label{eq:qsoln}
q = -\frac{i}{\sqt\!} e^{-i \alpha} \sqrt{r + r^3} \quad,
\qquad
\qt = \frac{i}{\sqt\!} e^{i \alpha} \frac{r^1 + i r^2}{\sqrt{r + r^3}}
\quad.
\end{equation}
This choice of $\alpha$ is convenient when we impose our gauge
condition: $q=-ip$ ($p>0$) corresponds to $\alpha=0$.  The effect of
an ordinary gauge transformation $q_i \to e^{-2 i \lambda} q_i$ is
$\alpha \to \alpha + 2 \lambda$; $\rvec$ remains gauge invariant.
This leads to a natural gauge invariant combination of real scalars
\begin{equation*}
\kappa \equiv \gamma + \alpha \quad.
\end{equation*}
This new scalar $\kappa$ is a good choice for the coordinate $r^4$ of
the Kaluza-Klein monopole: gauge variant coordinate fields have more
complicated supersymmetry transformations.

The constraints also contain information on the auxiliary fields.  The
constraint involving $A_\pm$ is the most notable of these:
\begin{equation}
\label{eq:Aeom}
\begin{aligned}
A_\pm & = \frac{1}{g^2 + 2 r} \left( - g^2 \partial_\pm \gamma
    + r \ovec \cdot \partial_\pm \rvec + 2 r \partial_\pm \alpha
    + 2 (\bar{\psi}_\pm \psi_\pm
         - \bar{\tilde{\psi}}_\pm \tilde{\psi}_\pm) \right)\\
 & = \frac{1}{g^2 H} \left( \partial_\pm \kappa
    + \frac{1}{2} \ovec \cdot \partial_\pm \rvec
    + \frac{1}{r} (\bar{\psi}_\pm \psi_\pm
                   - \bar{\tilde{\psi}}_\pm \tilde{\psi}_\pm) \right)
  - \partial_\pm \gamma
\quad.\end{aligned}
\end{equation}
This equation uses the (smeared) harmonic function $H = H(r)$
introduced in section~\ref{sec:MonopoleReview} and the target space
vector $\ovec$ defined (implicitly) by:
\begin{equation}
\label{eq:omegaDef}
i \left( q^\dag \dmu q - q \dmu q^\dag
            - \qt^\dag \dmu \qt
            + \qt \dmu \qt^\dag \right)
 = \frac{r^1 \dmu r^2 - r^2 \dmu r^1}{r+r^3} + 2 r \dmu \alpha
 \equiv r \ovec \cdot \dmu \rvec + 2 r \dmu \alpha \quad.
\end{equation}
In the first equality we have used Eq.~\eqref{eq:qsoln}, but the final
expression holds for redefinitions of $\alpha$ as described below.
The explicit form of $\ovec$ shown here is simply the Cartesian form
of the monopole gauge field mentioned in
section~\ref{sec:MonopoleReview}: $\omega_r = \omega_\vartheta=0$,
$\omega_\varphi = 1 - \cos \vartheta$.  While $\ovec$ does not
naturally change under gauge transformations, $\alpha$ does: $\delta(2
r \dmu \alpha) = 4 r \nabla \lambda \cdot \dmu \rvec$.  If we redefine
$\alpha$ and $\ovec$ so that $\alpha=0$ again holds, $\ovec$ changes
by $\delta \omega_\mu = 4 \nabla_\mu \lambda$, a target space gauge
transformation.

We can at last write the low energy action in components.  After
applying the constraints, the action finally reads as follows:
\begin{equation}
\label{eq:KLComponents}
\begin{split}
\mathcal{L} & = 
  - \frac{1}{2} H |\dmu \rvec|^2
  - \frac{1}{2} H^{-1} \left( \dmu \kappa
                  + \frac{1}{2} \ovec \cdot \dmu \rvec \right)^2\\
 & \quad \mbox{}
   + i H \bigl( 
            \bar{\chi}_+ \partial_- \chi_+ +
            \bar{\tilde{\chi}}_+ \partial_- \tilde{\chi}_+ +
	    \bar{\chi}_- \partial_+ \chi_- +
            \bar{\tilde{\chi}}_- \partial_+ \tilde{\chi}_-
   \bigr)\\
 & \quad \mbox{}
   - \frac{1}{4 H |r|^3} \left( \partial_- \kappa
       + \frac{1}{2} \ovec \cdot \partial_- \rvec \right)
     r^m \chisigchi{m}{+}
   - \frac{1}{4 H |r|^3} \left( \partial_+ \kappa
       + \frac{1}{2} \ovec \cdot \partial_+ \rvec \right)
     r^m \chisigchi{m}{-}\\
 & \quad \mbox{} +  \frac{1}{4 |r|^3} \epsilon_{mnp}
   \chisigchi{m}{+}\: r^n \partial_- r^p
   +  \frac{1}{4 |r|^3} \epsilon_{mnp}
   \chisigchi{m}{-}\: r^n \partial_+ r^p
   \\
 & \quad \mbox{}
  - \frac{3}{4 g^2 H |r|^5}
    r^m \chisigchi{m}{+}\: r^n \chisigchi{n}{-}
  + \frac{1}{4 g^2 H |r|^3}
    \chisigchi{m}{+}\: \chisigchi{m}{-}
 \quad.
\end{split}
\end{equation}
Here, $H$ is the harmonic function $H(r)$ defined in
section~\ref{sec:MonopoleReview}, and the scalar fields $\gamma$ and
$\alpha$ have everywhere combined into $\kappa$.  The vector
combination $\chisigchi{m}{\pm}$ was defined in
Eq.~\eqref{eq:chiSqrVector}.

In deriving this action, we have dropped total derivative terms, but
the term $\mathcal{L}_{\text{top.}}$ remains significant:
\begin{equation}
\label{eq:LtopDyon}
\mathcal{L}_{\text{top.}} = \eumn \dmu (\theta A_\nu)
 = \theta \eumn \dmu A_\nu + \eumn \dmu \theta A_\nu \quad,
\end{equation}
where $A_\nu$ should be replaced by its equation of motion in
Eq.~\eqref{eq:Aeom}.  The bosonic part of $\dmu \gamma + A_\mu$ is
precisely the large gauge transformation 1-form $\form{\Lambda}$
introduced in section~\ref{sec:MonopoleReview}, so the first of these
terms has the form of the dyonic $B$-field of Taub-NUT: $\form{B} =
-\theta d\form{\Lambda}$.  (The identification of this term in the
action as a $B$-field follows from Eq.~\eqref{eq:nonlinsigmacompts}
below.) Thus, although $\theta$ loses its geometrical meaning after
T-duality, it is still significant as the dyonic coordinate of the
Kaluza-Klein monopole.

The second term is harder to interpret, as the action does not treat
$\theta$ as a dynamical field; if $\theta$ is constant, this term
vanishes.  It would be interesting to find an action which correctly
encoded independent dynamics for a geometrical coordinate ($\kappa$,
here) and its dual coordinate ($\theta$), but no such formalism is
currently known.

\subsubsection{Real superfields and the nonlinear sigma model action}

To find the geometric meaning of these results, we can rewrite the
action as a supersymmetric nonlinear sigma model in terms of real
$\calN = (1,1)$ superfields whose scalar parts are the coordinate
fields.  We use conventions in which the $\calN = (1,1)$
supercoordinates are pure imaginary, $\bar{\theta}^\alpha = -
\theta^\alpha$, so the component expansion of a real superfield $R =
R^\dag$ is
\begin{equation*}
R = A + \sqt \tp \Omega_+ + \sqt \tm \Omega_- + i \tp \tm F \quad.
\end{equation*}
Here, $A$ and $F$ are real scalars and $\Omega$ is a real spinor.

To extract real superfields from our chiral superfields $\Psi$ and
$\Gamma$, we simply take their real and imaginary parts and impose the
pure imaginary condition on $\theta^\alpha$ (essentially setting the
real part of our $\calN = (2,2)$ $\theta^\alpha$s to zero).  So, for
instance, $\Psi_1 \equiv \left( \Psi + \Psi^\dag \right) / \sqt = r^1
+ \dotsb$ and $-g^2 \Gamma_1 \equiv -g^2 \left( \Gamma + \Gamma^\dag
\right) / \sqt = r^3 + \dotsb$.  From these coordinate superfields, we
can read off the real fermions $\Omega^m$ naturally associated to each
of the coordinates $r^m$ and to $\gamma$:
\begin{align}
\label{eq:OmegaDefs}
\Omega^m_\pm & = \left( \frac{\chi_\pm + \bar{\chi}_\pm}{\sqt\!},
                 \qquad
                 - i \frac{\chi_\pm - \bar{\chi}_\pm}{\sqt\!},
                 \qquad
		 i \frac{\tilde{\chi}_\pm - \bar{\tilde{\chi}}_\pm}
                  {\sqt\!}
\right),
& \Omega^\gamma_\pm & = 
		 \frac{\tilde{\chi}_\pm + \bar{\tilde{\chi}}_\pm}
                  {\sqt g^2}
\quad.
\end{align}
(In the $H$-monopole case, the $\Omega^m_\pm$ are identical because
the coordinates $r^m$ are not affected by T-duality, but
$\Omega^\theta_\pm = \mp g^2 \Omega^\gamma_\pm = \mp (\tilde{\chi}_\pm
+ \bar{\tilde{\chi}}_\pm)/\sqt$.)  The vector combination
$\chisigchi{m}{\pm}$ can be expressed in terms of these real fermion
fields:
\begin{equation}
\label{eq:omegasqrvec}
\chisigchi{m}{\pm} = i \Bigl(
\Omega^2_\pm \Omega^3_\pm + g^2 \Omega^1_\pm \Omega^\gamma_\pm\;,
 \quad
\Omega^3_\pm \Omega^1_\pm + g^2 \Omega^2_\pm \Omega^\gamma_\pm\;,
 \quad
\Omega^1_\pm \Omega^2_\pm + g^2 \Omega^3_\pm \Omega^\gamma_\pm
 \Bigr) \quad.
\end{equation}

The final step required before we can write the action in real
superfield form is to find the appropriate fermionic partner for the
gauge invariant coordinate $\kappa = \gamma + \alpha$.  We have
already found $\gamma$'s partner $\Omega^\gamma_\pm$ above, but
finding the fermionic partner for $\alpha$ is more subtle.  It can be
derived (up to an unimportant constant offset) by taking the
supersymmetry variation of Eq.~\eqref{eq:omegaDef} and solving for
$\delta_\xi \alpha$, giving the result
\begin{equation*}
\label{eq:alphavari}
\delta_\xi \alpha = \sqt \xi^\alpha \frac{1}{2 r}
  \left( g^2 \Omega^\gamma_\alpha
    - r \ovec \cdot \Ovec_\alpha \right) \quad.
\end{equation*}
This combines neatly with $\delta_\xi \gamma = \sqt \xi^\alpha
\Omega^\gamma_\alpha$ to yield $\delta_\xi \kappa \equiv \sqt
\xi^\alpha \Omega^4_\alpha$.  To substitute $\Omega^4_\alpha$ for
$\Omega^\gamma_\alpha$ in the expressions above, we can solve for the
latter:
\begin{equation}
\label{eq:omegaGamma4rel}
g^2 \Omega^\gamma_\alpha = H^{-1} \left( \Omega^4_\alpha
  + \frac{1}{2} \ovec \cdot \Ovec_\alpha \right)
\quad.
\end{equation}
When the real fermions are substituted in the kinetic terms of
Eq.~\eqref{eq:KLComponents}, the result is
\begin{multline}
\label{eq:fermionKinetic}
i H \bigl( \bar{\chi}_\pm \partial_\mp \chi_\pm +
           \bar{\tilde{\chi}}_\pm \partial_\mp \tilde{\chi}_\pm
 \bigr) =\\
 \frac{i}{2} H
   \left( \Ovec_\pm \cdot \partial_\mp \Ovec_\pm \right)
  + \frac{i}{2} H^{-1}
   \left( \Omega^4_\pm
     + \frac{1}{2} \ovec \cdot \Ovec_\pm \right)
   \left( \partial_\mp \Omega^4_\pm
     + \frac{1}{2} \ovec \cdot \partial_\mp \Ovec_\pm \right)
  + \orderO(\Omega \Omega \partial r)
\;.
\end{multline}
The final term results from $\partial_\mp \ovec$ and contributes to
the connection terms in the nonlinear sigma model action, which we
will not compute in detail.

In components, the general supersymmetric nonlinear sigma model action
is~\cite{Howe:1985pm}:
\begin{multline}
\label{eq:nonlinsigmacompts}
S = \frac{1}{2 \pi} \int d^2x \biggl[ 
  - \frac{1}{2} g_{m n} \dmu \phi^m \dumu \phi^n
  - \frac{1}{2} B_{m n} \eumn \dmu \phi^m \dnu \phi^n \\
  \mbox{}
  + \frac{i}{2}  g_{m n} \Omega_-^m D_+ \Omega_-^n
  + \frac{i}{2}  g_{m n} \Omega_+^m D_- \Omega_+^n
  + \frac{1}{4} R_{m n p q}
      \Omega_+^m \Omega_+^n \Omega_-^p \Omega_-^q
  \biggr] \quad.
\end{multline}
Here, $D_\pm \equiv D^\pm_0 \pm D^\pm_1$, where $D^\pm_\mu$ is the
covariant derivative defined with positive or negative torsion,
respectively.  $R_{mnpq}$ is the Riemann tensor defined with positive
torsion.  (Our conventions differ somewhat from those
of~\cite{Howe:1985pm}, including the overall normalization $1/2 \pi$.)

Comparing this to our component action in Eq.~\eqref{eq:KLComponents}
with the substitutions from Eqs.~\eqref{eq:omegasqrvec}
and~\eqref{eq:fermionKinetic}, we see that the form of the kinetic
terms agrees and that the metric is precisely that for the
Kaluza-Klein monopole given in Eq.~\eqref{eq:KKmonmetric}.  The
Riemann tensor components extracted from the 4-fermion terms agree
with those computed from the metric as well.

\section{Worldsheet instanton corrections}
\label{sec:Instantons}

Worldsheet instantons in the $H$-monopole gauged linear sigma model
have been analyzed by Tong~\cite{Tong:2002rq}.  As shown above, the
gauged linear sigma model for the T-dual Kaluza-Klein monopole is very
similar, so we will closely follow Tong's approach in this section.
When the arguments and calculations are identical in both cases (up to
differences in conventions), we will simply cite his results.

In carrying out the calculation we will use the language of classical
vacua with fixed values of the moduli. Of course, strictly speaking,
there is no such thing in $1+1$ dimensions as there is no symmetry
breaking by the Mermin-Wagner-Coleman theorem \cite{Mermin:1966fe,
Coleman:1973ci}. As in previous treatments, we work in the framework
of the Born-Oppenheimer approximation, where fast or high-momentum
modes are integrated out to give a low-energy description in terms of
a quantum corrected moduli space.

\subsection{The classical action for instanton sectors}

In gauged linear sigma models, worldsheet instantons correspond to
vortices of the gauge field.  We count these by
\begin{equation*}
k = -\frac{1}{2 \pi} \int F_{1 2} \quad.
\end{equation*}
These instantons include not only any holomorphic worldsheet
embeddings apparent from the theories' low energy nonlinear sigma
model limits but also what Witten~\cite{Witten:1993yc} calls
``point-like instantons''.  As neither the flat target space of the
$H$-monopole nor the Taub-NUT target space of a single Kaluza-Klein
monopole contain holomorphic two-cycles, it is these point-like
instantons that are relevant here.

The first step in finding the instanton action is to identify a
classical solution to represent each instanton sector $k \neq 0$.  Our
starting point is the bosonic component form of the gauged linear
sigma model action given in
Eqs.~\eqref{eq:LkinCompts}--\eqref{eq:LpotCompts}.  We must also
consider the effect of $\mathcal{L}_{\text{top.}} = \eumn \dmu (\theta
A_\nu) = -\theta F_{01} + \eumn \dmu \theta A_\nu$.  The first term is
topologically significant, and as discussed below
Eq.~\eqref{eq:LtopDyon} the impact of the second term is unclear
because the action does not treat $\theta$ as a dynamical field.  In
the $H$-monopole case, $\theta$ is constant in the $g \to 0$ limit
taken below, and we will make that assumption here.

To begin the calculation, we choose a specific classical vacuum for
our instanton solution to approach at large distance.  All vacua must
satisfy $\phi = \sigma = 0$, and we can use the $SU(2)$ R-symmetry to
set $\qt = 0$ without loss of generality.  The vacuum conditions in
this case are $r^1 = r^2 = 0$ and $|q|^2 = r^3 \equiv \zeta$, where
we define $\zeta$ as a constant parameterizing the vacuum.

Of course, the Mermin-Wagner-Coleman theorem implies that $\zeta=r^3$
cannot be a true modulus, and indeed there are no finite action
solutions of the equations of motion satisfying these vortex boundary
conditions.  This difficulty can be overcome by an analogue of the
``constrained instantons'' procedure~\cite{Affleck:1980mp}: we perform
our calculations in a limit of the parameters of the theory in which
appropriate BPS solutions exist, and then rely on supersymmetry to
protect the results as we return to general parameter values.  For
both the $H$-monopole and Kaluza-Klein monopole, an appropriate limit
is to take the Taub-NUT radius $g \to 0$.  This procedure is justified
by the final result of the calculation, in which instanton corrections
are finite even in the strict $g \to 0$ limit.

The next step is to identify the significant bosonic variations about
this chosen vacuum.  Only variations that could affect the gauge field
are relevant; others will merely increase the total Euclidean action.
We can exclude variations in $\phi$ and $\sigma$ from the start, and
variations in $\qt$, $r^1$, and $r^2$ are related by R-symmetry to
variations of $q$ and $r^3$ and will not reduce the action.  After
Wick rotating to Euclidean space (with $x^2 \equiv i x^0$) the
remaining action is
\begin{equation*}
S_{\text{E}} = \frac{i}{2 \pi} \int d^2x \left[
  \frac{F_{12}^2}{2 e^2} 
  + \frac{1}{2 g^2} \! \left( \dmu r^3 \right)^2
  + \frac{g^2}{2}   \! \left( \dmu \gamma + A_\mu \right)^2
  + \left| \mathcal{D}_\mu q \right|^2
  + \frac{e^2}{2} \! \left( |q|^2 - r^3 \right)^2 
  + i \theta F_{12}
  \right] \quad. 
\end{equation*}
When $g \to 0$, the $\gamma$ kinetic term drops out of the action
entirely.  On the other hand, variations of $r^3$ away from $\zeta$
are frozen out when $g \to 0$, even when $|q|^2 \neq \zeta$.

Thus, the relevant action for the instanton calculation is
\begin{equation*}
S = \frac{i}{2 \pi} \int d^2x \left[
  \frac{1}{2 e^2} F_{12}^2
  + \left| \mathcal{D}_\mu q \right|^2
  + \frac{e^2}{2} \left( |q|^2 - \zeta \right)^2 
  + i \theta F_{12}
  \right] \quad.
\end{equation*}
This is precisely the same action as in the $H$-monopole case: the
abelian Higgs model action at critical coupling plus a $\theta$ term.
Completing the square gives
\begin{equation*}
S = \frac{i}{2 \pi} \int d^2x \left[
  \frac{1}{2 e^2} \left( F_{12} \mp e^2 (|q|^2 - \zeta) \right)^2
  + \left| \mathcal{D}_1 q \pm i \mathcal{D}_2 q \right|^2
  + \left( \mp \zeta + i \theta \right) F_{12}
  \right] \quad.
\end{equation*}
The first two terms are strictly non-negative, so the minimal action
occurs when they are zero.  This provides a set of first order
Bogomol'nyi equations for the vortex solution:
\begin{equation*}
\label{eq:Bogomolnyi}
F_{12} = \pm e^2 (|q|^2 - \zeta)
 \qquad \text{and} \qquad
\mathcal{D}_{\bar{z}} q = 0 \quad
 (\text{or lower sign:\ } \mathcal{D}_{z} q = 0) \quad.
\end{equation*}
When integrated, the third term is proportional to the instanton
number $k$.  Choosing the $\pm$ sign to give the tightest lower bound
on the real part of the action (the top sign is preferred when $k>0$),
we find that the action when the Bogomol'nyi equations are satisfied
is
\begin{equation}
\label{eq:VortexAction}
S_k = |k| \zeta - i k \theta \quad,
\end{equation}
where we have defined $S_k = -i S$ in the given instanton sector (so
the path integral factor $e^{i S}$ becomes $e^{-S_k}$).

\subsection{The instanton sum and measure}

The sum over instanton configurations has two parts: a discrete sum
over sectors $k$ (each represented by a solution $\{A_\mu^{(k)},
q^{(k)}\}$ of the Bogomol'nyi equations) and an integral over zero
modes.  (Tong argues that the contributions of bosonic and fermionic
non-zero modes cancel in the present case.)  To evaluate this
integral, we must find the proper measure by identifying the bosonic
and fermionic zero modes of the solution.  We must also identify any
corrections to the instanton action that depend on the zero modes and
find the long distance behavior of the zero mode solutions themselves.
As the action here matches the $H$-monopole case, this section simply
summarizes the results of~\cite{Tong:2002rq} except as noted.

The first step is to identify the proper measure for the bosonic zero
mode integral.  We begin from the linearized Bogomol'nyi equations,
which together with a gauge fixing condition can be written as a
bosonic Dirac equation.  For $k>0$,
\begin{equation*}
\label{eq:BosDirac}
\Delta
  \begin{pmatrix} \delta A_{\bar{z}}\\ \delta q \end{pmatrix} = 0
  \quad,
\qquad \text{where} \quad
\Delta \equiv \begin{pmatrix}
  \frac{2 i}{e^2} \partial & q^\dag\\
  -q & i \overline{\mathcal{D}}
  \end{pmatrix} \quad.
\end{equation*}
Erick Weinberg~\cite{Weinberg:1979er,Weinberg:1981eu} used index
theory to show that these equations have $2|k|$ normalizable, linearly
independent zero mode solutions.  These form the multi-vortex moduli
space $\mathcal{M}_k$, which decomposes as $\mathcal{M}_k = \Reals^2
\times \tilde{\mathcal{M}}_k$.  The coordinates $X^\mu$ on $\Reals^2$
are Goldstone modes encoding the center of mass of the vortices; for
$k>0$ and $\mu = 1,2$ the corresponding linearized fields after gauge
fixing are
\begin{equation}
\label{eq:BosCMZeroModes}
\left\{ \delta_\mu A_\nu = F_{\mu \nu},
  \qquad
        \delta_\mu q = \mathcal{D}_\mu q \right\} \quad.
\end{equation}
The coordinates $Y^p$ on $\tilde{\mathcal{M}}_k$ ($p = 1, \dotsc,
2(k-1)$) encode the relative vortex positions.

The metric on $\mathcal{M}_k$ is defined by the overlap of the zero
modes~\cite{Samols:1991ne}; the proper overlap integral emerges from a
standard gauge-fixed zero mode calculation~\cite{Bernard:1979qt}.
This can be computed explicitly for the Goldstone modes above, for
which we find $g_{\mu \nu} = \zeta |k| \delta_{\mu \nu}$.  The metric
$\tilde{g}_{pq}$ on $\tilde{\mathcal{M}}_k$ remains unknown for any
$|k|>1$.  The bosonic zero mode integral is
\begin{equation}
\label{eq:BosonicZMMeasure}
\int d\mu_B
  = \int d^2X \prod_{p=1}^{2(|k|-1)} dY^p\:
    \frac{\sqrt{\det g}}{(2 \pi)^{|k|}}
  = \frac{\zeta |k|}{2 \pi} \int d^2X \prod_{p=1}^{2(|k|-1)} dY^p \:
    \frac{\sqrt{\det \tilde{g}}}{(2 \pi)^{|k|-1}} \quad.
\end{equation}

Next, we require the corresponding measure for the fermionic zero
modes.  There are $4k$ of these, related to the $2k$ bosonic zero
modes by the unbroken supersymmetries.  The superpartners of the
$X^\mu$ are Goldstino modes from the broken supersymmetries.  Two of
these result from the breaking of our explicit $\mathcal{N}=(2,2)$
supersymmetry and are parameterized by the Grassmann variables
$\alpha_1$ and $\alpha_2$, while the other two are related to these by
R-symmetry and are parameterized by $\tilde{\alpha}_1$ and
$\tilde{\alpha}_2$.  Explicitly, these are pairs of fermion fields
which are zero modes of the operator $\Delta$ or its complex conjugate
$\Delta^*$ (neither $\Delta^\dag$ nor $\Delta^T$ have any zero modes).
For $k>0$:
\begin{equation}
\label{eq:FermiCMZeroModes}
\begin{aligned}
\tfrac{i}{\sqt\!} \bar{\lambda}_+
    & = \tfrac{i}{2} \alpha_1 F_{12}\;, 
  & -\tfrac{i}{\sqt\!} \lambda_-
    & = -\tfrac{i}{2} \alpha_2 F_{12}\;, 
  & -\tfrac{1}{\sqt\!} \tilde{\lambda}_+
    & = \tfrac{i}{2} \tilde{\alpha}_1 F_{12}\;, 
  & -\tfrac{1}{\sqt\!} \bar{\tilde{\lambda}}_-
    & = -\tfrac{i}{2} \tilde{\alpha}_2 F_{12}\\
\psi_- & = \alpha_1 \mathcal{D} q\;, 
  & \bar{\psi}_+ & = \alpha_2 \overline{\mathcal{D}} q^\dag\;, 
  & \bar{\tilde{\psi}}_- & = \tilde{\alpha}_1 \mathcal{D} q\;, 
  & \tilde{\psi}_+ & = \tilde{\alpha}_2 \overline{\mathcal{D}} q^\dag
\end{aligned}
\quad.
\end{equation}
These results differ somewhat from those given in~\cite{Tong:2002rq}
for this action, and we show that our $\psi$ and $\tilde{\psi}$ zero
modes lead to a simpler expression for the four-fermion correlation
function when $|k|>1$.  As in the bosonic case, no explicit form is
known for the fermion zero mode partners of the relative coordinates
$Y^p$; we parameterize them by $\beta^p$ and $\tilde{\beta}^p$.

The overlap integrals that define the fermionic moduli space metric
$g$ arise from a zero mode calculation analogous to the bosonic case.
The result is the same apart from a shift in normalization, which can
be thought of as a change of zero mode basis in
Eq.~\eqref{eq:BosCMZeroModes} from $\mu = 1,2$ to $\mu = z, \bar{z}$
to match Eq.~\eqref{eq:FermiCMZeroModes}.  The measure for the fermion
zero mode integral is
\begin{equation}
\label{eq:FermionicZMMeasure}
\int d\mu_F
  = \int d^2\alpha\,d^2\tilde{\alpha} \prod_{p=1}^{2(|k|-1)} d\beta^p\,
    d\tilde{\beta}^p \frac{1}{\det g}
  = \left( \frac{2}{\zeta |k|} \right)^2
    \int d^2\alpha\,d^2\tilde{\alpha} \prod_{p=1}^{2(|k|-1)} d\beta^p\,
    d\tilde{\beta}^p \frac{1}{\det \tilde{g}}
\quad.
\end{equation}

While the zero modes found above each solve the linearized equations
of motion, they may interfere with each other when integrated up to
solutions of the full system.  This results in a four-fermion
contribution to the action by relative fermion zero
modes~\cite{Dorey:1997ij,Dorey:1997tr}:
\begin{equation}
\label{eq:FourFermiAction}
S_{\text{4-fermi}} = \frac{1}{4} \tilde{R}_{pqrs}
  \beta^p \beta^q \tilde{\beta}^r \tilde{\beta}^s \quad.
\end{equation}
Here, $\tilde{R}$ is the Riemann tensor on the relative vortex moduli
space $\tilde{\mathcal{M}}_k$.

Finally, we need to know the explicit long-distance limit of the
Goldstino mode solutions.  The long-distance limit of the bosonic
field $q$ was found in~\cite{deVega:1976mi}.  Using polar coordinates
on the Euclidean worldsheet, $z = \rho e^{i \vartheta}$, the solution
for a $k$-vortex solution centered at the origin as $\rho \to \infty$
is
\begin{equation*}
\label{eq:VortexSolnLimit}
|q|^2 \to \zeta \left(
  1 - l_k(Y^p,\vartheta) \sqrt{\frac{2 \pi L}{\rho}}\, e^{-\rho/L}
  \right) \quad,
\end{equation*}
where the characteristic vortex length scale is $L = (2 e^2
\zeta)^{-1/2}$.  The functions $l_k(Y^p,\vartheta)$ are unknown except
for the numerical constant $l_1 = 8^{1/4}$~\cite{Tong:2002rq}.  The
phase of $q$ in the vortex solution is important: in our conventions,
$q = |q| e^{i k \vartheta}$.  We also need to know the corresponding
$\rho \to \infty$ limit of the gauge field, $A_{\bar{z}} \to
\frac{i}{2} e^{i \vartheta} (-k/\rho + l_k(Y^p,\vartheta)
\sqrt{\pi/2L\rho}\; e^{-\rho/L})$, which together with this $q$
satisfies $\overline{\mathcal{D}}q=0$ to the given order in $\rho$.

From these results we can find the long-distance behavior of the
Goldstino mode $\psi_-$; the others will have either the
same profile or its conjugate.  For $k>0$,
\begin{equation}
\label{eq:FermionZMLimit}
\psi_- = \alpha_1 \mathcal{D} q
       \to \alpha_1 \sqrt{\zeta}\; l_k(Y^p, \vartheta) \:
  e^{i (k-1) \vartheta} \sqrt{\frac{\pi}{2 L \rho}}\; e^{-\rho/L}
\quad.
\end{equation}
The final square root and exponential will be denoted $S_F(X)$ below,
as they give the asymptotic behavior of the diagonal component of a
Dirac fermion propagator with mass $1/L$.

\subsection{Instanton corrections to the geometry}

We can now assemble these results to compute the instanton
contribution to the $\psi^4$ correlation function.  This will
correspond to a modified four-fermion vertex for $\psi$ in the low
energy effective action.  As these modifications must still be of the
geometric form in Eq.~\eqref{eq:nonlinsigmacompts}, they can be
interpreted as correcting the Riemann curvature tensor for the
monopole geometry.  The four insertions must be able to absorb all
four fermionic zero mode integrals, so the only non-vanishing set of
insertions in the $k$-instanton sector when $k>0$ is
\begin{equation*}
\begin{split}
G_4^{(k)}(x_1,x_2,x_3,x_4) & = \left. \left< \bar{\psi}_+(x_1) \psi_-(x_2)
  \tilde{\psi}_+(x_3) \bar{\tilde{\psi}}_-(x_4) \right>
   \right|_{k\text{-instanton}}\\
 & = \int d\mu_B d\mu_F \left[ \bar{\psi}_+(x_1) \psi_-(x_2)
  \tilde{\psi}_+(x_3) \bar{\tilde{\psi}}_-(x_4)
  e^{-S_k - S_{\text{4-fermi}}} \right] \quad.
\end{split}
\end{equation*}
If $k<0$, the conjugate holds.  The components of this expression can
be found in Eqs.~\eqref{eq:VortexAction}, \eqref{eq:BosonicZMMeasure},
\eqref{eq:FermionicZMMeasure}, \eqref{eq:FourFermiAction},
and~\eqref{eq:FermionZMLimit}, and together they give
\begin{multline*}
G_4^{(k)}(x_1,x_2,x_3,x_4)
 = \frac{1}{(2 \pi)^{d/2}} \frac{2 \zeta}{\pi |k|}
  e^{-|k| \zeta + i k \theta} \\
 \times \int d^2X \prod_{p=1}^{d}
   \left( dY^p d\beta^p d\tilde{\beta}^p \right)
  \frac{l_k^4(Y^p, \vartheta)}{\sqrt{\det \tilde{g}}}
  e^{-\frac{1}{4} \tilde{R}_{pqrs} \beta^p \beta^q
   \tilde{\beta}^r \tilde{\beta}^s}
  \prod_{i=1}^{4} S_F(X - x_i) \quad.
\end{multline*}
Here, $d=2(|k|-1)$ is the dimension of the relative moduli space.  The
worldsheet position appears here only in the propagator terms (which
we trust only when the $|X-x_i|$ are large) and in the $\vartheta$
dependence of $l_k^4(Y^p, \vartheta)$, which characterize the vortex
solution falloff at large distance.  We expand any such dependence on
$\vartheta$ as a Taylor series and proceed using only the term without
higher derivative corrections: the $\vartheta$-averaged value of
$l_k^4(Y^p, \vartheta)$.  We can then separate out all terms involving
the relative moduli space into a function
$\nu(\tilde{\mathcal{M}}_k)$:
\begin{equation*}
G_4^{(k)}(x_1,x_2,x_3,x_4)
 = \frac{2 \zeta}{\pi |k|}
   e^{-|k| \zeta + i k \theta} \nu(\tilde{\mathcal{M}}_k)
   \int d^2X \prod_{i=1}^{4} S_F(X - x_i) \quad.
\end{equation*}
Explicitly, the function $\nu(\tilde{\mathcal{M}}_k)$ is
\begin{equation*}
\begin{split}
\nu(\tilde{\mathcal{M}}_k)
 & =  \frac{1}{(2 \pi)^{d/2}} \int \left[ \prod_{p=1}^{d}
   \left( dY^p d\beta^p d\tilde{\beta}^p \right)
  \frac{e^{-\frac{1}{4} \tilde{R}_{pqrs} \beta^p \beta^q
   \tilde{\beta}^r \tilde{\beta}^s}}{\sqrt{\det \tilde{g}}}
  \frac{1}{2 \pi} \int d \vartheta \,
  l_k^4(Y^p, \vartheta) \right] \\
 & = \frac{1}{(-8 \pi)^{d/2} (d/2)!}
  \int \prod_{p=1}^{d} dY^p \, \sqrt{\det \tilde{g}} \,
  \epsilon^{p_1 p_2 \ldots p_d} \epsilon^{q_1 q_2 \ldots q_d}
  \tilde{R}_{p_1 p_2 q_1 q_2} 
  \cdots \tilde{R}_{p_{d-1} p_d q_{d-1} q_d}\\
 & \hspace{9em} \times
 \frac{1}{2 \pi} \int d \vartheta \:
  l_k^4(Y^p, \vartheta) \quad.
\end{split}
\end{equation*}
Note that $\epsilon^{\cdots}$ is the usual contravariant volume
element whose non-zero components have magnitude $1/\sqrt{\det
\tilde{g}}$.  For $|k|=1$ no calculation is necessary: $\nu = l_1^4$.
For higher $|k|$, this expression differs from the same result
in~\cite{Tong:2002rq}: the phases of the four fermion zero modes
cancel out, so there is no exponential of $i \vartheta$ weighting the
exponential falloff function.  Thus, this is simply the integral of
the Euler form over the $k$-vortex moduli space, weighted by the
average exponential falloff.

The final modification to the four-fermion term in the low energy
effective action is found from the sum over instanton sectors,
$-\sum_k G_4^{(k)}$ (the minus sign appears after Wick rotating back
to a Lorentzian worldsheet).  Restoring $r$ in place of $\zeta$, we
obtain (up to a possible unimportant numerical factor)
\begin{align*}
\delta \mathcal{L}_{\text{eff}} & =
  - \sum_{k=1}^\infty \frac{2 r}{\pi |k|}
  \nu(\tilde{\mathcal{M}}_k) e^{-k r}
  \left[
  e^{i k \theta}
    \bar{\psi}_+ \psi_- \tilde{\psi}_+ \bar{\tilde{\psi}}_-
  + e^{-i k \theta}
    \bar{\psi}_- \psi_+ \tilde{\psi}_- \bar{\tilde{\psi}}_+
  \right] \nonumber \displaybreak[0] \\
 & = - \sum_{k=1}^\infty \frac{1}{2 \pi |k| r}
  \nu(\tilde{\mathcal{M}}_k) e^{-k r}
  \left[
  e^{i k \theta}
    \tilde{\chi}_+ \bar{\tilde{\chi}}_- \chi_+ \bar{\chi}_-
  + e^{-i k \theta}
    \tilde{\chi}_- \bar{\tilde{\chi}}_+ \chi_- \bar{\chi}_+
  \right] \nonumber \displaybreak[0] \\
 & = - \sum_{k=1}^\infty \frac{1}{8 \pi |k| r}
  \nu(\tilde{\mathcal{M}}_k) e^{-k r} \Bigl[
{ \renewcommand{\arraystretch}{1.5}
\begin{array}[t]{@{}r@{}l@{}c@{}l@{}}
    & e^{i k \theta} & 
    (\Omega^1_+ + i \Omega^2_+) &\left(\Omega^3_+ + i H^{-1}
    (\Omega^{4}_+ + \frac{1}{2} \ovec \cdot \Ovec_+ ) \right) \\
    & \hfill \mbox{} \times \mbox{} &
    (\Omega^1_- - i \Omega^2_-) & \left(\Omega^3_- - i H^{-1}
    (\Omega^{4}_- + \frac{1}{2} \ovec \cdot \Ovec_- ) \right) \\
    \mbox{} + \mbox{} & e^{-i k \theta} &
    (\Omega^1_+ - i \Omega^2_+) & \left(\Omega^3_+ - i H^{-1}
    (\Omega^{4}_+ + \frac{1}{2} \ovec \cdot \Ovec_+ ) \right) \\
    & \hfill \mbox{} \times \mbox{} &
    (\Omega^1_- + i \Omega^2_-) & \left(\Omega^3_- + i H^{-1}
    (\Omega^{4}_- + \frac{1}{2} \ovec \cdot \Ovec_- ) \right)
  \Bigr]
\end{array}
}
 \nonumber \displaybreak[0] \\
 & \xrightarrow{g \to 0}
   - \sum_{k=1}^\infty \frac{1}{8 \pi |k| r}
   \nu(\tilde{\mathcal{M}}_k) e^{-k r} \Bigl[
{ \renewcommand{\arraystretch}{1.8}
\begin{array}[t]{@{}r@{}l@{}c@{}l@{}}
    & e^{i k \theta} & 
    (\Omega^1_+ + i \Omega^2_+) \Omega^3_+
    & (\Omega^1_- - i \Omega^2_-) \Omega^3_- \\
    \mbox{} + \mbox{} & e^{-i k \theta} &
    (\Omega^1_+ - i \Omega^2_+) \Omega^3_+ 
    & (\Omega^1_- + i \Omega^2_-) \Omega^3_-
  \Bigr]
 \quad.
\end{array}
}
\end{align*}
As this is part of the low energy action, we have used
Eq.~\eqref{eq:comptConstraints} and our vacuum choice $\qt = 0$ in
deriving the second line and Eqs.~\eqref{eq:OmegaDefs}
and~\eqref{eq:omegaGamma4rel} to find the third.  In taking the final
limit, we have dropped terms involving $H^{-1} \propto g^2$ as our
calculation may have neglected terms of this same order.  By
comparison with Eq.~\eqref{eq:nonlinsigmacompts} (and after accounting
for the symmetries of the Riemann tensor) we can see that the net
coefficient of $\Omega^m_+ \Omega^n_+ \Omega^p_- \Omega^q_-$ here is
$\delta R_{mnpq}/(2 \pi)$.

This result can be extended to $r^1, r^2 \neq 0$ by R-symmetry
(corresponding to rotations in target space), but by construction it
will hold only in a particular large distance limit.  We have kept
only the leading term at large $r$ for each instanton sector, but we
sum over higher $|k|$ modes despite their exponential suppression.
Treating the small quantities $1/r$ and $e^{-r}$ independently in this
way is physically reasonable because the exponential terms have a
distinct origin as higher instanton sectors.

For the terms which survive the final $g \to 0$ limit, this result for
$\delta \mathcal{L}$ is identical to the $H$-monopole case: as noted
previously, the low energy component constraints in
Eq.~\eqref{eq:comptConstraints} and the real scalar superpartners
$\Omega^{1,2,3}_\pm$ are the same for both monopole solutions.  Thus,
the leading corrections to the components of the Kaluza-Klein monopole
Riemann tensor with indices 1, 2, and~3 must be the same as those for
the localized $H$-monopole.

To leading order in $1/r$, it can be verified that the Riemann tensor
with torsion is
\begin{equation*}
R_{mnpq} =  -\frac{1}{2} ( \partial_m \partial_p g_{nq}
      + \partial_n \partial_q g_{mp} - \partial_m \partial_q g_{np}
      - \partial_n \partial_p g_{mq} )
    - \frac{1}{2} ( \partial_{m} T_{npq} - \partial_{n} T_{mpq} )
 \quad.
\end{equation*}
This assumes that both $\Chris{m}{pq}$ and $\Tors{m}{pq}$ fall off at
least as $1/r$, which does hold in our case.  Applying these formulas
to the localized $H$-monopole geometry given in
section~\ref{sec:MonopoleReview}, we can (for instance) find the
curvature corrections to leading order in $1/r$ evaluated at
$r^1=r^2=0$ for comparison to the instanton result above:
\begin{equation*}
\label{eq:RiemannCorrections}
\begin{array}{r@{\eqsp}r@{\eqsp}r@{}c@{}l@{\eqsp}l}
\vspace{0.5em}
\delta R_{1313} & \delta R_{2323}
 & - & \displaystyle \frac{1}{4 r}
 & \displaystyle \sum_{k=1}^{\infty} k^2 e^{-kr}
  (e^{i k \theta} + e^{-i k \theta})
 & \displaystyle
 -\frac{1}{2 r} \sum_{k=1}^{\infty} k^2 e^{-kr} \cos (k \theta) \\
\delta R_{1323} & -\delta R_{2313}
 && \displaystyle \frac{i}{4 r}
 & \displaystyle \sum_{k=1}^{\infty} k^2 e^{-kr}
  (e^{i k \theta} - e^{-i k \theta})
 & \displaystyle
 -\frac{1}{2 r} \sum_{k=1}^{\infty} k^2 e^{-kr} \sin (k \theta)
 \quad.
\end{array}
\end{equation*}
Naturally, we need not limit ourselves to $r^1=r^2=0$ in general.  The
physical meaning of these corrections is different in the two cases:
in the Kaluza-Klein monopole, $\theta$ is now the dyonic coordinate
rather than a part of the geometry.  In interpreting these results, it
is useful to recall that $R_{pqmn} = R_{mnpq}|_{T \to -T}$ whenever
the torsion is a closed form (which always holds in string theory,
where $\form{T} = -d\form{B}$).  This allows us to recognize that the
terms in the first line must come entirely from corrections to the
metric while the terms in the second line must come from an
instanton-induced torsion.

While many geometries would have this same limit at large distance,
the connection to the localized $H$-monopole's Riemann tensor
corrections suggests that the metric corrections are the same:
\begin{equation}
\delta g_{11} = \delta g_{22} = \delta g_{33}
 = \frac{1}{2 r} \sum_{k=1}^\infty e^{-k r}
  \left( e^{i k \theta} + e^{-i k \theta} \right) \quad.
\end{equation}
This corresponds to a correction to the harmonic function $H$ in those
metric components to the form in Eq.~\eqref{eq:Hlocalized}:
\begin{equation*}
H = \frac{1}{g^2} + \frac{1}{2r}
  \sum_{k=-\infty}^{\infty} e^{-|k|r + i k \theta}
  = \frac{1}{g^2} + \frac{1}{2r}
  \frac{\sinh r}{\cosh r - \cos \theta} \quad.
\end{equation*}
It seems likely that in the full corrected solution (beyond our $g \to
0$ limit), the harmonic function is modified in this way in all
components of the Kaluza-Klein monopole metric.  Whether there are
additional corrections is unclear.

The instanton corrections also generate a torsion, in contrast to the
usual Kaluza-Klein monopole solution.  The only component that is
non-zero to first order in $1/r$ in the $g \to 0$ limit is
\begin{equation}
\label{eq:torsioncorrection}
T_{123} = -H_{123}
 = -\frac{1}{r} \sum_{k=1}^\infty k e^{-k r} \sin(k \theta)
 \quad.
\end{equation}
This may be written in terms of the localized harmonic function
$H(r,\theta)$ as $H_{123} = \partial_\theta H$.

\section{Interpretation and conclusions}
\label{sec:InterpretConclude}

\subsection{Winding space localization of the Kaluza-Klein monopole}

To understand these corrections, we must return to the conjecture
of~\cite{Gregory:1997te} that the proper Kaluza-Klein monopole in
string theory should have some sort of ``throat'' behavior, just as
the NS5-brane does.  (Strictly speaking, a throat is only present for
higher monopole charge, but there are hints of it even in the poorly
understood unit charge case.)  In particular, that paper suggested
that just as the $H$-monopole throat can be probed by strings with
momentum along $\theta$, the Kaluza-Klein monopole throat could be
probed by strings winding around $\kappa$.  Meanwhile, the geometrical
isometry along $\kappa$ remains unbroken.

As part of that work, \cite{Gregory:1997te}~studied the behavior of
winding strings in the Kaluza-Klein dyon geometry.  That analysis
showed that although strings can unwind from the $\kappa$ circle in
various ways, a generalization of the winding charge remains
conserved.  Each change in string winding number is offset by a finite
shift in the ``velocity'' $\dot{\beta}(t)$, where $\beta(t)$ is the
dyonic coordinate introduced in Eq.~\eqref{eq:dyoncoord}.
Intuitively, $\beta$ is the coordinate on ``winding space'', and
string winding charge is equivalent to momentum in $\beta$.

As seen in Eq.~\eqref{eq:LtopDyon}, after T-duality from the
$H$-monopole the role of this dyonic coordinate is played by $\theta$:
``momentum space'' has become ``winding space''.  The corrections
found above give strong evidence that the conjectured localization and
throat behavior do appear.  The modified harmonic function $H(r,
\theta)$ has the same form that described a throat in the
$H$-monopole, but it now appears only for a special value of the
winding space coordinate $\theta$ rather than at a special point
around the geometrical circle.  And as expected from duality, the
resulting torsion provides a mechanism for this structure to couple to
winding strings.

Our interpretation of the corrections to the Kaluza-Klein monopole
solution differs somewhat from that of~\cite{Gregory:1997te}.  That
paper viewed this winding space localization as a coherent state of
classical string winding modes, in analogy with an interpretation of
the localized $H$-monopole as a coherent state of string momentum
modes.  Intuitively, this picture is exactly right: the localized
monopole solutions can be expanded in Fourier modes that carry the
correct conserved charges.  However, the classical solutions for
strings with momentum and winding are known, and superpositions of
those solutions with the weights predicted by~\cite{Gregory:1997te} do
not give the proper correction terms on either side of T-duality.

We expect this monopole to leave some supersymmetry unbroken just as
the NS5-brane does, but at first this seems impossible.  One of the
conditions for unbroken supersymmetry is that the dilatino variation
vanish:
\begin{equation*}
\left( \gamma^m \partial_m \Phi
 - \frac{1}{6} \gamma^{mnp} H_{mnp} \right) \xi
 \stackrel{?}{=} 0 \quad.
\end{equation*}
For non-trivial solutions $\xi$ to exist, the $\gamma$ matrices must
factor out to leave a projection operator $1 \pm \gamma^5$.  This is
possible only if the coefficients of $\gamma^m$ and $\gamma^m
\gamma^5$ are equal in magnitude for each $m$.  In particular, for
$m=4$ this condition requires $|\partial_4 \Phi| = |H_{123}|$ (with
tangent space indices).  As we have found that $H_{123} \neq 0$, this
holds only if $\partial_4 \Phi \neq 0$, but we have seen no physics
that would break the $\kappa$ isometry.  (Changing from curved to
tangent space indices does not solve the problem.)

The resolution to this puzzle is that the usual supergravity
approximation is only expected to hold when the radius $g$ of the
$\kappa$ circle is large and momentum states are light.  Because we
have performed the instanton calculation in the limit of small $g$,
the proper light degrees of freedom are instead the winding states and
a different low-energy theory must apply.  T-duality suggests that it
should formally agree with the supergravity description of the
$H$-monopole at large radius, involving the dyonic coordinate $\theta$
rather than the geometrical coordinate $\kappa$.  In particular, it
seems likely that the relevant part of the equation for unbroken
supersymmetry in this case will be
\begin{equation}
\label{eq:newSUSY}
\left( \gamma^\theta \partial_\theta \Phi
 - \gamma^{123} H_{123} + \dotsb \right) \xi = 0 \quad.
\end{equation}
Here, $\gamma^\theta$ denotes the matrix $\gamma^4$ from the
$H$-monopole, and in the $g \to 0$ limit, $\gamma^{123}$ agrees with
that case as well.  As noted below Eq.~\eqref{eq:torsioncorrection},
$H_{123} = \partial_\theta H$ (with curved space indices), so if
Eq.~\eqref{eq:newSUSY} is valid the dilaton should be $e^{\Phi} =
H(r,\theta)$ just as for the $H$-monopole.  This would lead to a
throat behavior at a particular value of $\theta$, which we expect to
persist even for finite $g$.

\subsection{Conclusions and open questions}

While the Kaluza-Klein monopole geometry is well known, its familiar
form does not correspond to the full solution in string theory.  The
usual form is ``smeared'' in winding space, and worldsheet instanton
effects lead to its localization there.  The corrections involved are
very similar to those that localize the smeared $H$-monopole, but they
explicitly depend on the Kaluza-Klein monopole's dyonic coordinate
rather than the geometrical coordinate on the circle.

This work leaves a number of interesting questions unanswered.
Perhaps the most basic of these is the exact form of the corrected
geometry itself: our calculation was carried out in the strict $g \to
0$ limit and only to leading order in $1/r$.  A better understanding
supergravity when string winding states are light could be a helpful
step in that direction, and would have importance in its own right.

Another natural extension of this work is to look for similar
corrections to other objects in the duality web.  Kaluza-Klein
monopoles also appear in M-theory, and this work would seem to suggest
that those solutions receive similar corrections from membrane
instantons.  It could also be instructive to study the case of higher
monopole charge: true throat behavior does not emerge in the
$H$-monopole until the charge is greater than one, and the same is
presumably true for Kaluza-Klein monopoles as well.

Finally, it remains clear that the winding space coordinate $\theta$
appears in the action in a fundamentally different way than the
geometrical coordinate $\kappa$.  While differences are certainly
expected, it is odd to find that the action does not appear to specify
the dynamics of $\theta$ at all.  It would be valuable to develop a
more symmetric description of geometrical coordinates and their duals.
Such a formalism could be important in finding an appropriately
generalized supergravity theory as well.

\begin{acknowledgments}
We would like to thank David Kutasov, Itai Seggev, and Michael Seifert
for helpful discussions.  This work was supported in part by NSF Grant
No. PHY-0204608.  SJ also received support from an ARCS Foundation
scholarship.
\end{acknowledgments}

\bibliography{monopoles}

\end{document}